# Boiling Heat Transfer on Superhydrophilic, Superhydrophobic, and Superbiphilic Surfaces

*Amy Rachel Betz[1*], James Jenkins[2], Chang-Jin "CJ" Kim[2], Daniel Attinger[3]*
[1] Kansas State University, Manhattan, KS
[2] University of California, Los Angeles, CA
[3] Iowa State University, Ames, IA

* Contact Information:
3002 Rathbone Hall
Manhattan, KS 66506
arbetz@ksu.edu
785-532-2647

# Abstract

With recent advances in micro- and nanofabrication, superhydrophilic and superhydrophobic surfaces have been developed. The statics and dynamics of fluids on these surfaces have been well characterized. However, few investigations have been made into the potential of these surfaces to control and enhance other transport phenomena. In this article, we characterize pool boiling on surfaces with wettabilities varied from superhydrophobic to superhydrophilic, and provide nucleation measurements. The most interesting result of our measurements is that the largest heat transfer coefficients are reached not on surfaces with spatially uniform wettability, but on *biphilic* surfaces, which juxtapose hydrophilic and hydrophobic regions. We develop an analytical model that describes how biphilic surfaces effectively manage the vapor and liquid transport, delaying critical heat flux and maximizing the heat transfer coefficient. Finally, we manufacture and test the first *superbiphilic surfaces* (juxtaposing superhydrophobic and superhydrophilic regions), which show exceptional performance in pool boiling, combining high critical heat fluxes over 100 W/cm$^2$ with very high heat transfer coefficients, over 100 kW/m$^2$K.

**Keywords:** Superhydrophobic, superhydrophilic, biphilic, enhanced heat transfer, pool boiling, nucleation

## Nomenclature

| | | |
|---|---|---|
| $A_{in}$ | = | area of influence (m$^2$) |
| $c_p$ | = | specific heat (J/kgK) |
| CHF | = | critical heat flux (W/m$^2$) |
| $d$ | = | diameter (m) |
| $f$ | = | frequency (Hz) |
| $g$ | = | acceleration due to gravity (m/s$^2$) |
| $k$ | = | thermal conductivity (W/mK) |
| HTC | = | heat transfer coefficient (W/m$^2$K) |
| $n_a^{'}$ | = | active nucleation site density (sites/m$^2$) |
| $p$ | = | pitch (m) |
| $q''$ | = | heat flux (W/m$^2$) |
| SBPi | = | superbiphilic |
| SHPi | = | superhydrophilic |
| SHPo | = | superhydrophobic |
| $T$ | = | temperature (°C or K) |
| $\Delta T$ | = | T-T$_{sat}$, superheat (K) |

*Greek Symbols*

| | | |
|---|---|---|
| $\gamma$ | = | surface tension (N/m) |
| $\theta$ | = | wetting angle (°) |
| $\rho$ | = | density (kg/m$^3$) |

*Subscripts*
  *con*    =   contact
   *d*     =   departure
   *l*      =   liquid
  *sat*   =   saturation
   *v*     =   vapor

## 1. Introduction

A surface is called superhydrophilic (SHPi) if the apparent contact angle of water on the surface in air is close to zero, which induces spontaneous spreading. The high affinity of SHPi surfaces for water enhances capillary water transport [1, 2], prevents dropwise condensation or fogging [3, 4], and facilitates boiling [5]. A surface is called superhydrophobic (SHPo) if the apparent contact angle of water on the surface in air is larger than 150°. These surfaces are inspired by natural structures such as the lotus leaf and have a wealth of technical applications. SHPo surfaces self-clean [6], enhance condensation [7], mitigate frost buildup [8, 9], and reduce hydrodynamic drag [1, 10]. Typically, fabrication of SHPi or SHPo surfaces requires engineering a water-attracting or repelling surface to have a severe roughness on the sub-millimeter scale, which increases or decreases the true contact area with water, respectively. In recent reviews, enhanced liquid-vapor phase change was also described as a potential application of SHPi [11] and SHPo [12] surfaces. This potential enhancement is important for convective heat transfer technologies since convection associated with multiphase flow delivers the highest heat transfer coefficients, typically one order of magnitude higher than single-phase forced convection, and two orders of magnitude higher than single-phase natural convection [13]. Few multiphase heat transfer measurements however have been made on SHPi or SHPo surfaces, with the exception of experiments involving single nucleation [14] or condensation [15].

The high heat transfer rates delivered by boiling are needed in industrial applications such as thermal generation of electricity, metallurgy, electronics cooling, and food processing. While *flow boiling* describes the boiling of liquids forced to move along hot solid surfaces, *pool boiling*, the mode studied here, describes a fluid heated on a hot surface and transported by buoyancy [16]. Two parameters measure the pool boiling performance. First, the heat transfer coefficient (HTC) is the ratio of the heat flux ($q''$) to the difference between the surface temperature and the boiling temperature of the fluid ($\Delta T$) or HTC=$q''/\Delta T$. The HTC describes the thermodynamic efficiency of the boiling exchange. Second, the critical heat flux (CHF) is the highest heat flux that a surface can exchange with a boiling fluid before the individual bubbles merge into a vapor layer that insulates the surface from the liquid. In the regime with $q''<CHF$, HTC typically increases with $q''$, as the solid surface interacts with an increased number of liquid and vapor pockets, maximizing the opportunity to transfer heat and mass across the liquid wedges of the multiple wetting lines. At CHF, the HTC is drastically reduced, which induces a significant and often destructive surface temperature increase [17-19] called dry-out.

To date, two main strategies have been used to enhance the performance of surfaces for pool boiling. The first strategy enhances the performance at low heat fluxes, in the isolated bubble regime, by promoting nucleation and enhancing HTC [16]. This is made by either reducing the surface wettability [14, 20-26] or by modifying the surface topology, via e.g. surface roughening, etching of cavities [20, 27, 28], or microporous coatings [29, 30]. The second strategy enhances the performance at high heat fluxes, in the regime of slugs and columns [16], which results in an enhanced CHF. This is made by improving liquid transport, typically by increasing surface wettability [31, 32], which also sharpens wetting angles and steepens thermal gradients [31]. Wettability can be enhanced by increasing the roughness of a hydrophilic surface [33] at the sub-millimeter scale. Note that some micro- and nanostructuring processes used to increase wettability come with the benefit of randomly distributed microcavities and defects [5, 17, 30, 34], which also facilitate nucleation.

In this study, we fabricate surfaces with engineered wettability as shown in Figure 1. We measure for the first time the density of active nucleation sites on SHPi and SHPo surfaces, an important input parameter needed for numerical simulations of boiling on such surfaces [35]. Our modeling and pool boiling measurements also show how the wettability of a surface, as well as the juxtaposition of regions of different wettabilities, control and enhance boiling heat transfer.

## 2. Materials and Methods

## 2.1 Design and manufacturing of enhanced surfaces for pool boiling

### 2.1.1 Surface design and fabrication:

Six types of surfaces are fabricated for this study, as shown in Figure 1. The four types in the top row have spatially uniform wettabilities (hydrophilic, hydrophobic, SHPi, SHPo). The two types shown in the bottom row of Figure 1 juxtapose hydrophilic and hydrophobic regions: this design induces a concurrent affinity for water and for water vapor, a *quality that we name biphilic*. Nature has examples of biphilic surfaces that enhance multiphase heat transfer. The biphilic wings of the Namib desert beetle optimize its water intake [7]; while hydrophilic regions of the wing help condensation, the hydrophobic regions guide the liquid to its mouth. Few biphilic surfaces have been fabricated [25, 28, 36], but they all have been shown to significantly enhance boiling heat transfer. In 1965, the first biphilic surface by Hummel [25], who sprayed hydrophobic polymer drops onto a steel surface, showed a HTC 2 to 7 times higher than the bare steel surface. Biphilic surfaces were recently fabricated using microlithographic techniques [21, 28, 36]. The microfabricated biphilic surfaces by Betz et al. [28], shown on the bottom left of Figure 1, exhibited not only HTCs 100% larger but also a CHF 65% larger than a hydrophilic surface.

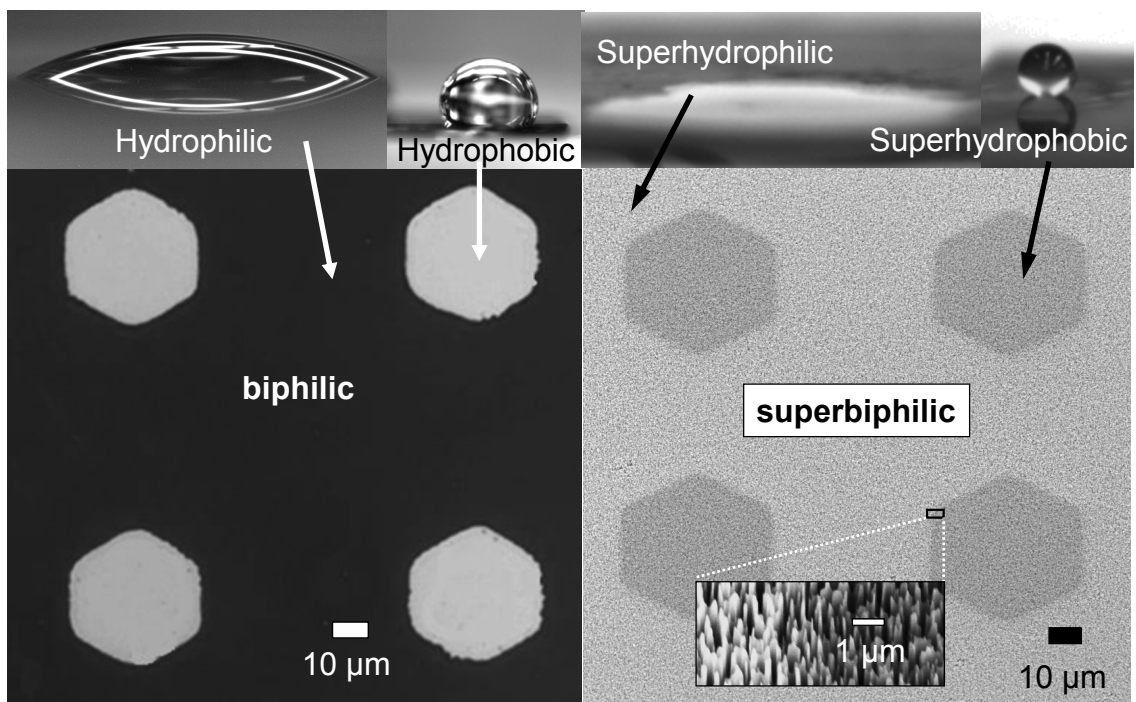

**Figure 1. Six types of surfaces are considered in this study. The first four types (top row, from left to right) have spatially uniform wettability: an oxidized silicon hydrophilic surface (7-30º wetting angle, as shown by the imaged water drop), a fluoropolymer-coated hydrophobic silicon surface (110-120º), a SHPi surface (0º), and a SHPo surface (150-165º). The fifth type is a biphilic surface, bottom left, which juxtaposes hydrophilic and hydrophobic regions, as indicated by the arrows. The sixth type is a superbiphilic (SBPi) surface, bottom right, which juxtaposes SHPi and SHPo regions. That SEM picture appears grainy because of the surface nanostructuring, and a magnified view is in the inset.**

Pushing the *biphilic* concept to more extreme values of wettability, we have also manufactured superbiphilic (SBPi) surfaces. SBPi surfaces juxtapose SHPo and SHPi areas, as shown in the bottom right of Figure 1. These surfaces were manufactured on silicon wafers using a combination of random nanostructuring processes, microlithography, and thin hydrophobic polymer coating, as follows. The random nanostructures are made by a DRIE using the black silicon method [37]. Next, the entire surface is exposed to oxygen plasma in a RIE machine for 30 minutes to create a 30 nm silicon dioxide layer, rendering the surface SHPi. To obtain a microscale pattern of SHPo areas on the SHPi field, a photolithographic process was employed on the SHPo surfaces. Teflon® fluoropolymer is spun onto the entire surface and baked; this additional coating is thin –less than 100nm thick – and smooth – less than 5nm rough. It preserves the original random structures of the etched silicon, as can be seen in the inset of Figure 1. Photoresist with added surfactant (to aid wetting on the fluoropolymer surface) is spun onto the surface. The photoresist is patterned using photolithography. The exposed fluoropolymer is removed by oxygen plasma in an RIE machine for 3 minutes. Where the coated fluoropolymer is etched away, the underlying oxidized nanostructures, i.e., SHPi surface, are exposed. Figure 1 also characterizes the wettability of each surface, using visualization of 100 μL drops at ambient temperature.

### 2.1.2 Heater fabrication:

Thin film heaters made of indium tin oxide (ITO) are directly deposited on the reverse side of the silicon wafer used to create the SBPi surfaces. The surface has a thermally grown oxide layer for electrical passivation. First, ITO is sputtered onto the silicon wafer in a custom Angstrom deposition chamber. The heater geometry of 1 cm x 3 cm is obtained by using a polycarbonate shadow mask. A target resistivity of 50 ohms/square is used to determine the ITO thickness, which was typically 300 nm. Copper electrodes of 1 cm x 1 cm were thermally deposited onto each end of the ITO heater, also using a polycarbonate shadow mask, leaving a 1cm x 1 cm square of ITO exposed. The heater was electrically passivated by depositing a 50-100 nm layer of $SiO_2$ using a Semicore e-beam evaporator.

## 2.2 Pool boiling measurements

### 2.2.1 Fabrication of a test assembly:

A silicon wafer, fabricated as in the previous section, is placed on a Teflon® gasket that holds the wafer in place with the heater side up to prevent the surface with patterned wettability from being contaminated or scratched. Three braided wires are attached to each copper electrode using silver paint. A thin film thermocouple is placed over the center of the heater and attached using polyimide tape. A strip of silicon glue is spread along the edge of the wafer to form a wall around the wafer while allowing thermocouple and electrical wires access through the sides. The inside of the silicon glue barrier is filled with PDMS for thermal insulation. The PDMS is mixed vigorously just before pouring to ensure a maximum number of air bubbles in the mixture to lower the thermal conductivity. The test assembly (wafer piece + thermocouple + PDMS) is heated on a hot plate at 100 °C for at least 1 hour to cure the PDMS. The final thickness of the PDMS layer is 5 – 10 mm. Finally, the wires are connected to the power supply and the thermocouple is connected to the data acquisition device.

### 2.2.2 Pool boiling measurement:

The pool boiling setup is similar to the one used in our previous work [28]. A cubic pool is made from polycarbonate, with outer dimensions of 70x70x70 mm. On one side is a Pyrex window for visualization. The test assembly is placed in the pool with the pattern-side up. The test assembly is held in place by two Teflon® rods. The pool is filled with thoroughly degassed water. Two 100 W submerged cartridge heaters are placed in the pool and set to constant power to heat and maintain the pool at saturation temperature. After the pool has reached a steady temperature the heat flux applied to the heater is increased. Once a stable temperature is reached at a given heat flux the temperature is recorded for 300 measurements acquired at 1 Hz. Since there is some temperature fluctuation in the measurements, the temperature is recorded over time to make sure that this variance is periodic.

**2.2.3 Measurement uncertainties**

The maximum combined uncertainty on the heat flux was estimated as 3.5 % of the heat flux. This was caused by the combined measurement uncertainty on the heater area and the current and the voltage measurements. We also performed experiments to determine the heat lost through the PDMS insulation as a function of the heater temperature. This was done by exposing the wafer side of the test assembly to air (where convection is negligibly lower than in water), while maintaining the rest of the test assembly in the water pool at saturation temperature and applying various heat fluxes. We found that the heat lost through the insulation is a linear function of the heater temperature, corresponding to about 0.45W/K, and the reported values of the heat flux have been corrected for that loss. The maximum uncertainty on the superheat was estimated as ±1.5 K, due to the thermocouple uncertainty, temperature acquisition, and heater/wafer thickness measurement uncertainties. Due to the maximum thermocouple error of ±1.5 K, the uncertainty of the HTC can be greater than 100 % at superheat values lower than 1 K. This error decreases as the superheat increases and is less than 20 % of the HTC at superheats above 5 K and less than 10 % at superheats above 15 K.

## 2.3 Analytical modeling

To describe and explain how the thermal performance in the isolated bubble regime depends on wettability or on patterns of wettability, we develop an analytical model. The starting point is the micro-convection model of Mikic and Rohsenow [38], which assumes that rising bubbles act as intermittent pumps enhancing convective heat transfer:

$$\text{HTC} = 2(\pi k_l \rho_l c_{pl})^{1/2} n_a' d_d^2 f^{1/2}. \qquad (1)$$

In equation (1), the symbols $k_l$, $\rho_l$, and $c_{pl}$ represent the thermal conductivity, the density, and the specific heat capacity of the liquid phase, respectively. The HTC is expressed as the product of these material properties with the density of active nucleation sites $n_a'$, the square of the departure diameter of the bubble $d_d$, and the square root of the frequency of bubble departure $f$.

The first parameter needed to solve equation (1) is the density of active nucleation sites, $n_\alpha'$. This density depends on the geometry and chemistry of the surface [20, 39], and is best estimated experimentally. We provide such measurement in Section 3.1. In our model for HTC, we also consider that the surface will eventually become saturated with bubbles, constraining the maximum number of nucleation sites to $n_{a,\max}'$. This is achieved by assuming that each bubble is surrounded by an area of influence $A_{in} = 2\pi(d_d/2)^2$, from which liquid is drawn for the bubble growth [40].

The second parameter in equation (1), the departure diameter, is found by assuming that bubbles depart when buoyancy forces overcome surface tension forces. This mechanism best represents the reality at low heat fluxes, when convective shear forces are negligible. For a surface with uniform wettability, Figure 2a shows that the departure diameter and maximum contact diameter depend only on the wetting angle. While the maximum contact diameter increases monotonically, the departure diameter reaches its maximum around a wetting angle of 110º.

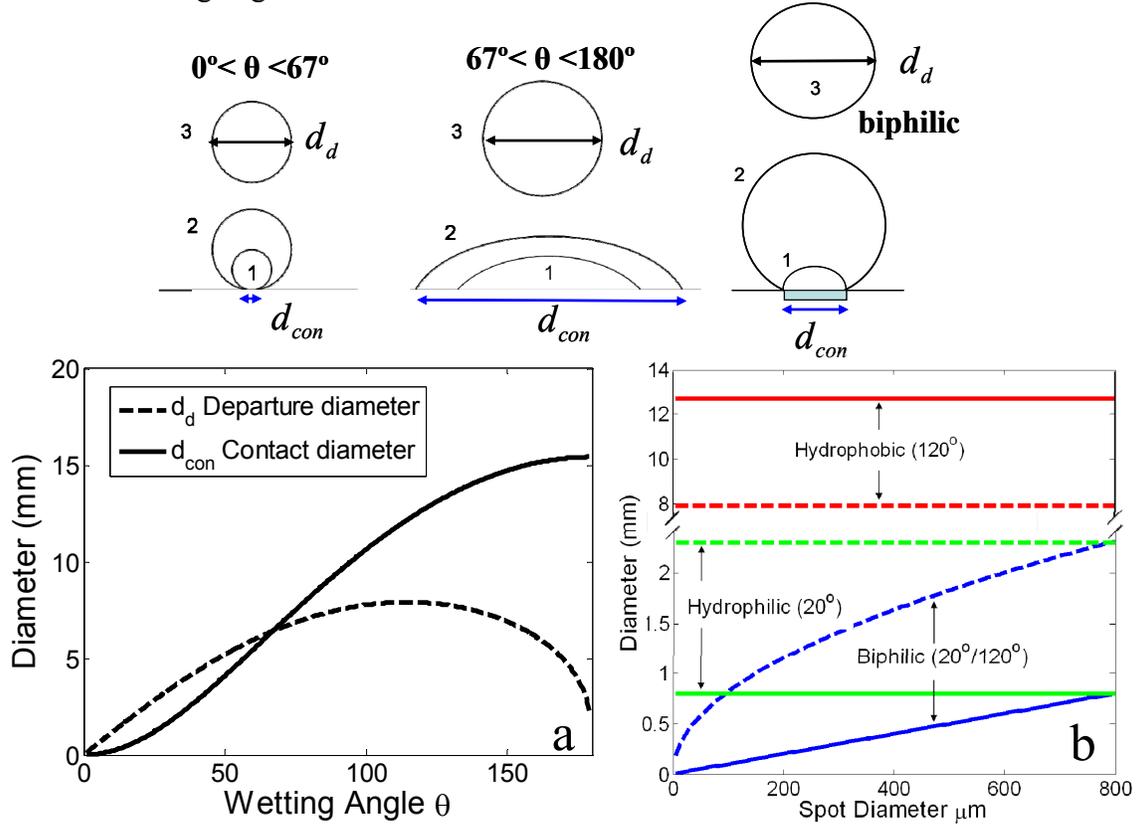

**Figure 2:** Model of bubble growth and departure based on wetting and buoyancy forces. The departure diameter $d_d$ of a vapor bubble and its maximum contact diameter $d_{con}$ depend on the wetting angle for a surface of uniform wettability (a), and also on the diameter of the hydrophobic spot on a biphilic surface (b). The cartoons above the plots illustrate the growth and departure of the bubble.

The last parameter of equation (1), the bubble frequency, is typically found in experiments to be inversely proportional to the departure diameter [41],

$$fd_d = C\left[\frac{\gamma g(\rho_l - \rho_v)}{\rho_l^2}\right]^{1/4}, \qquad (2)$$

where the coefficient $C$ is dependent on the working fluid of interest (for water, C=0.59).

## 3. Results and Discussion

### 3.1 Nucleation curves on surfaces with spatially uniform wettability

The surfaces in Figure 1 were characterized in the pool boiling setup described in Section 2.2.2, allowing for optical access and the measurement of heat flux vs. surface temperature. In Figure 3, high-speed visualization is used to characterize the incipience of boiling, i.e. the density of active nucleation sites as a function of the surface superheat

*ΔT*. The surfaces compared in Figure 3 are the four types with spatially uniform wettability, i.e. the hydrophilic, hydrophobic, SHPi and SHPo surfaces. The wetting and non-wetting surface demonstrated different behavior. For the hydrophilic and SHPi surfaces the number of nucleating bubbles and departing bubbles is the same. On these surfaces, the number of nucleation sites can be counted at any moment from the high-speed video. However, for the hydrophobic and SHPo surface, many bubbles nucleate and then quickly merge, resulting in a single bubble departing from the respective surface. For the number of nucleation sites on hydrophobic and SHPo surfaces, we report the number of nucleating bubbles, measured immediately after bubble departure.

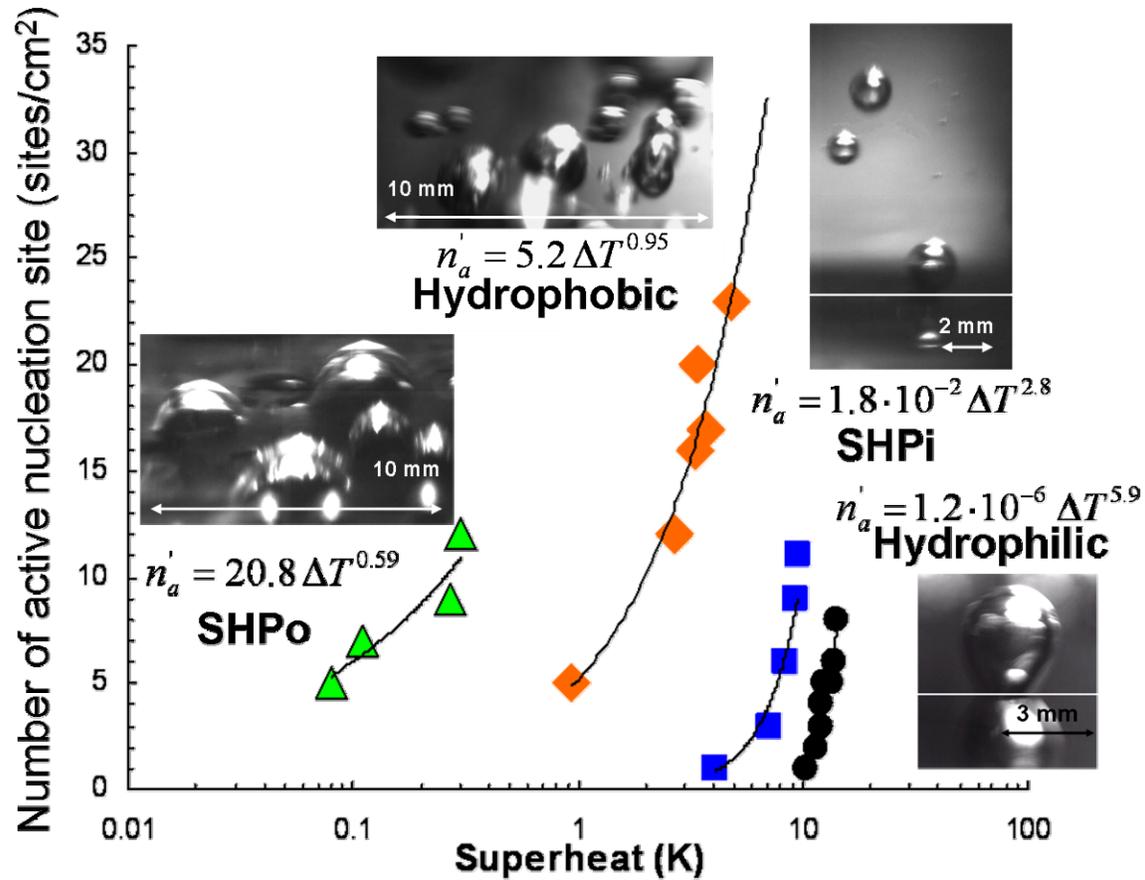

**Figure 3. Measured density of active nucleation sites for a hydrophilic, hydrophobic, SHPi and SHPo surface, as a function of the superheat. A power law fit is provided for modeling purposes, and visualization of the bubbles are provided, with a white line showing the solid-fluid interface.**

Figure 3 first shows that the SHPi surface (blue squares) nucleates at a surface superheat about three times lower than the hydrophilic surface (black circles). In previous work on the pool boiling performance of nano- and microstructured SHPi surfaces, the enhancement of HTC was indeed attributed to an increase in available nucleation sites [5, 34]. More significant information from Figure 3 is that a very strong nucleation enhancement is seen on hydrophobic and SHPo surfaces. Hydrophobic surfaces (orange diamonds) nucleate at values of superheat about one order of magnitude lower than hydrophilic surfaces (black circles), and SHPo surfaces (green triangles) nucleate at superheats another order of magnitude lower than hydrophobic surfaces (orange diamonds). Note that Figure 3 reports that SHPo surface nucleates at superheat temperatures significantly lower than the typical thermocouple measurement uncertainty of ±1.5 K. To best quantify such low superheat values, the test sample was heated in the pool boiling setup at saturation temperature until it reached thermal equilibrium. At that point, the temperature of the thermocouple attached the heater was recorded as $T_{sat}$. The heat flux was then slowly increased until nucleation was visible on the surface: at that point the thermocouple temperature $T_{nucleation}$ was measured. Since these two temperatures are very close for SHPo surfaces, the expected uncertainty is typically lower than that of a standard thermocouple measurement. Nevertheless, measurements of nucleation on SHPo surfaces call for more accurate temperature measurement methods, such as resistive temperature devices [42, 43] or arrays of thin film thermocouples [44]. Note that the very low superheat values measured in Figure 3 are compatible with classical nucleation theory, which predicts for smooth SHPo surfaces that the free energy needed to nucleate bubbles vanishes as $\theta \to \pi$ [16, 45].

These orders-of-magnitude enhancement of nucleation rates on hydrophobic and SHPo surfaces should drastically improve HTC in comparison with hydrophilic surfaces. To the best of our knowledge only one previous work [26] looked at nucleation on a SHPo surface, mostly qualitatively, and found that bubbles actually form at negative superheat values and that a vapor film covers the surface before any bubble departure. Indeed, the nucleation enhancement on hydrophobic and SHPo surfaces comes with the drawback that they reach CHF at low heat fluxes, in the range of 30 W/cm$^2$; this is due to their strong tendency to form an insulating vapor film, a phenomenon called the Leidenfrost effect.

## 3.2 Boiling curves on surfaces with uniform wettability

With the above estimations of $n_a^{'}$, $d_d$ and $f$, equation (1) is used to determine HTC as a function of ΔT for a surface of given, uniform wettability. Comparisons of the modeled HTC with the HTC measured in this work are provided in Figure 4a. The model and experiments agree well for the boiling behavior of hydrophobic and hydrophilic surfaces, in terms of the maximum HTC and the superheat needed to reach that maximum. Compared with the hydrophobic surface (orange diamonds), the hydrophilic surface (black circles) provides a lower HTC at a lower superheat but a higher HTC at a higher superheat.

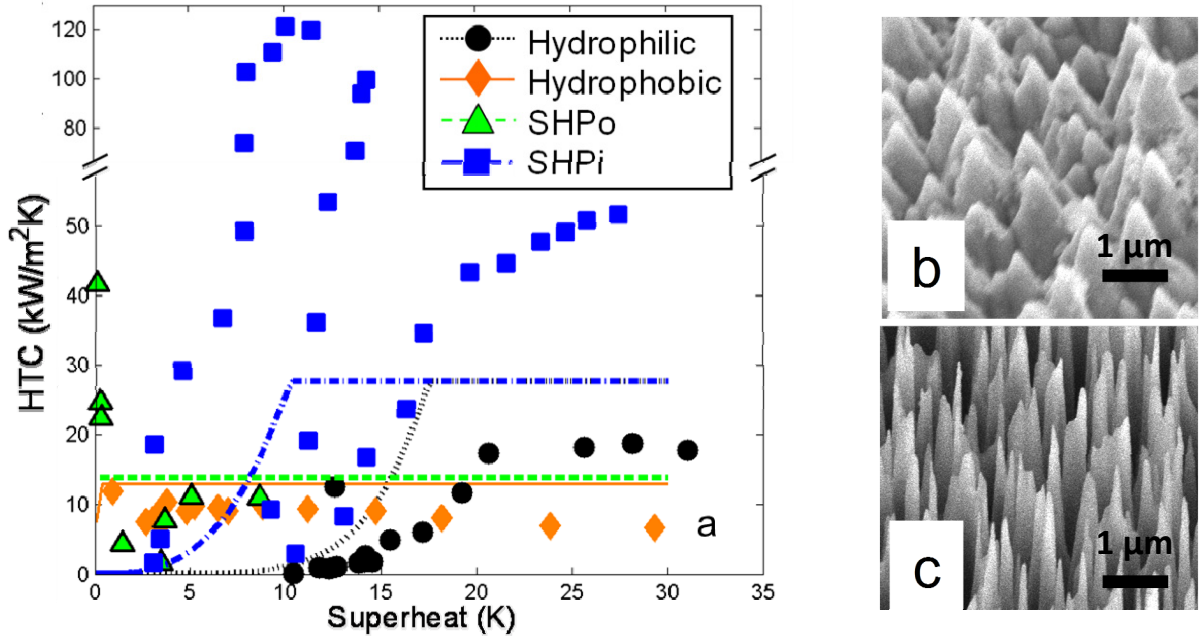

**Figure 4. (a) Boiling measurements (points) for a hydrophilic, hydrophobic, SHPi and SHPo surface, compared with our theoretical model (lines). Insets (b) and (c) show the variation in surface topology for two different SHPi surfaces.**

In Figure 4a, the hydrophobic surface features large HTC at low superheat, but this value stays constant at larger superheat because the large contact diameter of the bubbles (see Figure 2) limits the maximum number of active nucleation sites. The hydrophilic surface provides a higher HTC than the hydrophobic surface, albeit at a higher superheat. Those two findings suggest that the wettability that optimizes HTC is a function of the superheat at which the surface operates. In Figure 4a, the SHPi surfaces exhibit the highest measured HTC of all surfaces with spatially uniform wettability. These results confirm available heat transfer measurements on nano-engineered surfaces [5, 34]. The modeling results in Figure 4a consistently underpredict the very high HTC values obtained on SHPi surfaces, probably because the model does include all the physics, for instance the additional wicking created by the micro- and nano-roughness. Also, Figure 4a shows significant measurement noise for very low $\Delta T$ on SHPo surfaces (green triangles), because the typical thermocouple measures temperature with an uncertainty (±1.5K) on the same order as the low $\Delta T$ (see discussion on measurement uncertainties in Section 2.2.3). Significant HTC noise is also visible on the SHPi surfaces, which might be attributed to the random nature of the nanostructuring process used, such as peak density (0.8-3.8 peaks/$\mu m^2$), peak height (0.7-1.98 $\mu m$) and peak width at the base of the structure (0.3-1 $\mu m$), as visible in the two samples of SHPi surfaces in Figure 4b-c. Finally, it is worth mentioning that the wetting angle used in the model for SHPi surfaces is not 0°, which would correspond to bubble with null departure diameter (see Figure 4a), but 20°, which corresponds to the observed departure size of the bubbles measured on SHPi surfaces.

## 3.3 Boiling curves on biphilic surfaces, effect of wettability and topography

Next, we show that HTC can be further increased by revisiting an assumption underlying our measurements in Section 3.2. This assumption is that the boiling surface has a spatially uniform wettability. This assumption is questionable, since an *ideal* boiling surface has contradictory requirements on wettability: it requires hydrophobicity to promote nucleation and enhance HTC in the regime of isolated bubbles, and it requires hydrophilicity to maintain water transport to the hot surface in the regime slugs and columns, which results in a high CHF [31]. We propose to use the biphilic and superbiphilic surfaces of Section 2.1 to resolve this apparent contradiction and optimize heat transfer performance.

The cartoons and graphs in Figure 2 describe how a biphilic surface combines the advantages of both a hydrophobic surface (large bubble departure diameter $d_d$ and abundance of nucleation sites) and of a hydrophilic surface (contact diameter smaller than the departure diameter, which prevents the merging of adjacent bubbles). For the HTC of a biphilic surface, we modified equation (1), from Section 2.3, by assuming that the total heat flux is the sum of two heat fluxes, transferred in parallel across both hydrophobic and hydrophilic regions. This is expressed by equation (3) below.

$$\text{HTC} = 2(\pi k_l \rho_l c_{pl})^{0.5} \left[ (f^{0.5} d_d^2 n_a')_{hydrophilic} + (f_{spot}^{0.5} d_{d,spot}^2 n_a')_{hydrophobic} \right] \quad (3)$$

Results in Figure 2b show that the geometry of departing bubbles on a biphilic surface depends on the wettability contrast, and on the size of the hydrophobic spot. As a result, a biphilic surface offers more control than a surface with uniform wettability on the bubble nucleation, growth and detachment. This additional control might help enhance pool boiling performance. Note that we assume that the wetting line of the bubble advances until pinning occurs at the edge of the hydrophobic spot; also the number of active nucleation sites of the hydrophobic regions cannot exceed the number of hydrophobic spots.

In Figure 5a, the influence of the wettability contrast of biphilic surfaces on HTC is quantified theoretically and experimentally. The influence of this contrast is theoretically studied by varying the wetting angle on the hydrophilic regions (20°,7°,3°), while keeping the wettability on the hydrophobic regions constant. At lower values of superheat, the HTC is independent on the hydrophilic wetting angle, probably because most nucleation and boiling occur on the hydrophobic regions. At superheat values higher than 15 K, the HTC is increased when the wettability contrast is increased. Modeling results compare well with experimental results for wetting angles of 20° and 7°. These wetting angle values correspond respectively to a surface with an untreated thermally grown oxide layer and to a surface rinsed in a low concentration HF solution, which etches a thin layer of the oxide. We were not able to manufacture surfaces with 3° wetting angles, so only modeling results are shown. The agreement between experiments and modeling is good in terms of trends and absolute values.

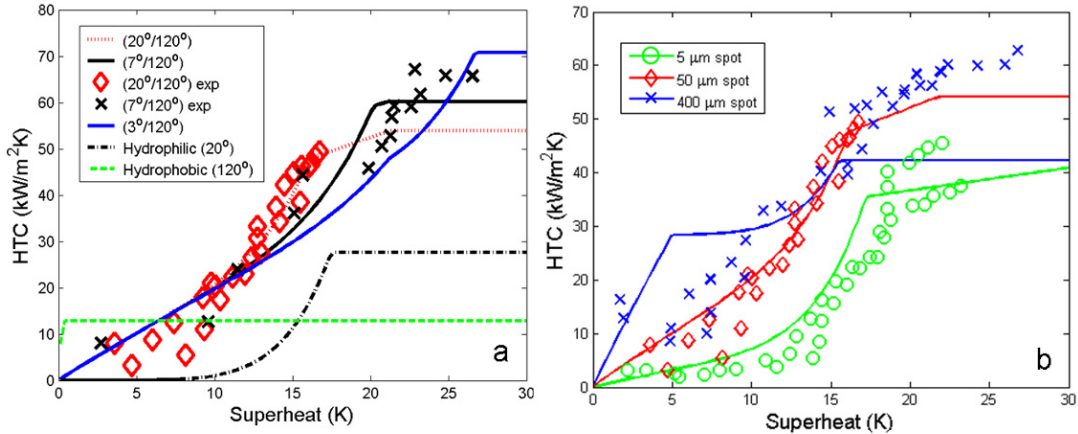

**Figure 5.** Boiling curves of biphilic surfaces with wettability contrast as in the legend, with solid curves denoting modeling and dotted curves, experiments: (a) Effect of the wettability contrast of biphilic surfaces (b) Effect of the spot size of the biphilic surface, for surfaces with wettability contrast of (20°/120°). All biphilic surfaces have a spot diameter/pitch ratio $d/p$ = 0.5. For comparison purposes, (a) also shows modeling results for surfaces with uniform wettability, with corresponding experimental results available in Figure 4 and not reproduced here for clarity.

Comparison with hydrophilic and hydrophobic surfaces, in Figure 5a, shows that the maximum HTC of the biphilic surface is two and four times larger than the HTC of the hydrophilic and hydrophobic surface, respectively. This result shows that *biphilic surface features larger HTC than surfaces with uniform wettability* because they combine the advantages of both hydrophilic and hydrophobic surfaces: like hydrophilic surfaces, biphilic surfaces generate bubbles with departure diameters larger than their contact diameters (shown in Figure 2b), thereby offering large HTC at large superheat; like hydrophobic surfaces, biphilic surfaces offer more nucleation sites at low $\Delta T$ than hydrophilic surfaces.

Theoretical and experimental curves in Figure 5b investigate the effect of the *topography* on HTC, by varying the diameter of the hydrophobic spots, at constant pitch to diameter ratio. Experiments and theory show that 5 μm spots induce lower HTCs than the 400 and 50 μm spots. The agreement between experiments and theory for the 50 and 5 μm spots is good, better at low superheat ($\Delta T$<15K) than at larger superheat ($\Delta T$>15K), probably because the modeling neglects shear forces. Regarding surfaces with 400 and 50 μm spots, the theoretical curves predict that at low superheat the 400 μm spots induce higher HTC, while at larger superheat, the 50 μm spots induce higher HTC. This can be explained by the fact that at low superheat the bubbles released from the 400 μm spots are larger, while at higher superheat the surface with 400 μm spots offers less nucleation sites than the surface with smaller spots. *This last finding suggests that there is no unique optimal surface topography, but rather that the optimum topography depends on operating conditions such as superheat.*

## 3.4 Boiling curves on superbiphilic surfaces

Considering that SHPo surfaces have the largest density of nucleation sites (see Figure 3), and that HTC increases with the wettability contrast on biphilic surfaces (Figure 5a), we designed and studied surfaces that juxtapose SHPi and SHPo areas. These surfaces, the first *superbiphilic* (SBPi) surfaces to the best of our knowledge, are fabricated as described in Section 2.1.1. Preliminary results in Figure 6 compares the thermal preformance of a SBPi surface with a biphilic surface of identical topography (50 μm spots with $d/p$ =0.5), with a hydrophilic surface, and with a SHPi surface. The SBPi surfaces reached HTC over 150 kW/m$^2$K, confirming the intuition that SBPi surfaces would reach the highest HTC. Compared to a smooth hydrophilic surface (SiO$_2$, contact angle 7º), the improvement in HTC in pool boiling is larger than one order of magnitude at low superheat (best shown from 5K to 10K) and about 300% for larger values of superheat. Note that the measured performance of the SBPi surfaces is higher than predicted by the analytical model, possibly because the model only accounts for effects of wettability contrast and not for capillary transport enhancement caused by the surface nanostructuring. The variation between samples may be attributed to the random nature of the nanostructuring process employed, see Section 2.1.1. The increase in HTC of the SBPi surfaces over SHPi surfaces is probably due to the increased availability of nucleation sites provided by the SHPo spots, and to the ability to geometrically control the distribution of nucleation sites. We found no significant influence of the shape (circles and hexagons tested) of the SHPo spots. Both the analytical modeling and the experiments show that SBPi surfaces outperform all other surfaces in the low superheat regime (ΔT<10K).

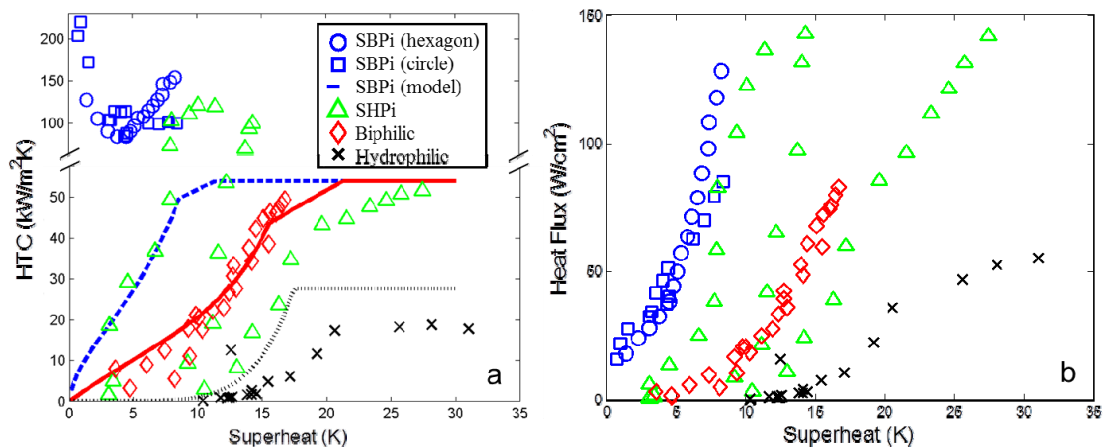

**Figure 6. Boiling curves comparing hydrophilic, biphilic, SHPi and SBPi surfaces, (a) is HTC vs. superheat and (b) shows the heat flux vs. superheat. All surfaces biphilic have the same topography, 50 μm spots with d/p =0.5. The wetting contrast of the biphilic surface is (20°/120°).**

The performance of biphilic and SBPI surfaces is best considered in Figure 7, which compares the CHF (a) and HTC (b) of biphilic and SBPi surfaces with a few state-of-the-art nanostructured surfaces made of silicon or copper nanowires [5, 34], and with the classical Rohsenow correlation for pool boiling of water on top of a smooth copper surface [46]. While SBPi surfaces have CHFs comparable to state-of-the-art nanostructured surfaces, SBPi surfaces however offer higher HTCs, enhanced by a factor up to three for low values of superheat. The values of HTC in Figure 7b are larger than 100kW/m$^2$K, and are the highest values reported to date in pool boiling of water on flat surfaces. Note the uncertainty at low heat flux, due to thermocouple measurement uncertainties (see Section 2.2.3). Considering Figure 7a-b together, one obtains a clearer idea of how SBPi surfaces enhance both the HTC and CHF: SBPi surfaces enhance HTC by facilitating nucleation on the hydrophobic spots, while the hydrophilic background prevents early CHF and allows for reaching high CHFs. However, it could be misleading to consider SBPi surfaces as a competitor of the many types of enhanced nanofabricated surfaces recently developed. Superbiphilicity is rather a topographic architecture of the surface that can be obtained using a variety of nanofabrication methods for specifically fabricating the juxtaposed SHPi and SHPo regions. In that sense, superbiphilicity is an additional tool to enhance the multiphase performance of nanofabricated surfaces.

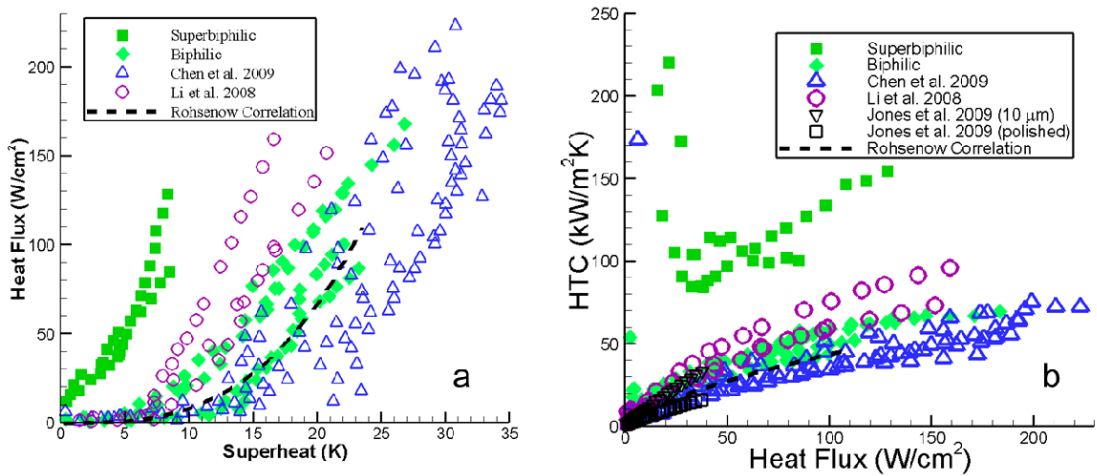

**Figure 7. (a) Heat flux vs. superheat for biphilic and SBPi surfaces and (b) is HTC vs. superheat. For comparison are recently published results with two state-of-the-art nanofabricated surfaces, [5, 34] the Rohsenow correlation for water on a smooth copper surface,[46] and two measurement of water on copper with different roughness [33].**

## 4. Conclusions

This study describes the design and fabrication of biphilic surfaces which juxtapose hydrophilic and hydrophobic regions. We show experimentally that these surfaces have higher performance in pool boiling than surfaces with spatially uniform wettability, in terms of critical heat flux and heat transfer coefficient. We show with an analytical modeling how the excellent boiling performance is due to the biphilicity of the surfaces: while the hydrophobic regions increase the availability of nucleation sites, the surrounding hydrophilic regions constrain the contact diameter of the growing bubbles, preventing the surface from being saturated with bubbles, i.e. delaying critical heat flux. The study also provides measurements of the density of active nucleation sites on superhydrophobic and superhydrophilic surfaces. Finally, we design and characterize the first superbiphilic (SBPi) surfaces, which juxtapose superhydrophobic and superhydrophilic regions. Heat transfer coefficients measured on SBPi surfaces are up to three times higher than on state-of-the-art nanostructured surfaces. Importantly, SBPi and biphilic surfaces are not a competitor to the various surface nanostructuring methods for heat transfer enhancement: *biphilicity* is a topographic architecture consisting of arranging wettability contrasts on a surface, the local wettability of which can be obtained using existing micro- and nanofabrication methods. Biphilic and SBPi surfaces are likely to improve a wide range of transport phenomena that involve moving wetting lines and capillary phenomena, from boiling to condensation. Future work will aim at a better understanding and control of multiphase flow on biphilic surfaces by means of, e.g., parametric studies on the surface topography. The long-term stability of these surfaces will also be characterized towards the development of technical applications.

**Acknowledgements:**
We are grateful for the help of the reviewers in improving the quality of the experiments and the manuscript.